\documentclass[aps,twocolumn,nofootinbib,nobibnotes]{revtex4}
\usepackage{epsfig}
\usepackage{dcolumn}
\usepackage{bm}

\usepackage{graphicx}
\usepackage{latexsym}

\def\be{\begin{equation}}
\def\ee{\end{equation}}
\def\bea{\begin{eqnarray}}
\def\eea{\end{eqnarray}}

\begin{document}

\title{Cosmological Evolution of Interacting Dark Energy Models  \\
with Mass Varying Neutrinos}
\author{Xiao-Jun Bi}
\affiliation{Key laboratory of particle astrophysics,
Institute of High
Energy Physics, Chinese Academy of Sciences, P.O. Box 918-3,
Beijing 100049, People's Republic of China}
\author{Bo Feng}
\author{Hong Li}
\author{Xinmin Zhang}
\affiliation{Theoretical Physics Division, Institute of High
Energy Physics, Chinese Academy of Sciences, P.O. Box 918-4,
Beijing 100049, People's Republic of China}

\begin{abstract}

In this paper we consider the cosmological implications of dark
energy models with a coupled system of a dynamical scalar field (the
quintessence) and the neutrinos. By detailed numerical calculations we
study the various possibilities on the evolution and the fates
of the universe in this class of models. Our results show that due
to the interaction with quintessence, neutrinos could be dominant
over the quintessence in the future universe, however would
eventually decay away.

\end{abstract}
\maketitle

\section{Introduction}

Recent data from type Ia supernova (SN Ia) \cite{pel} and cosmic
microwave background (CMB) radiation \cite{wmap} have provided
strong evidences for a spatially flat and accelerating
universe at the present time. In the context of
Friedmann-Robertson-Walker cosmology, this acceleration is attributed to
the domination of a new matter component with negative pressure,
dubbed dark energy \cite{tur}.
The simplest candidate for dark energy
seems to be a remnant small cosmological constant or vacuum energy with
$\rho \sim {( 2 \times 10^{-3}  eV )}^4$. This energy scale $\sim
10^{-3}$ eV is smaller than the known energy scales in particle physics
except that of the neutrino masses, which is
comparable to the scale of dark energy.

The dark energy could also be due to a dynamical component, such as a
canonical scalar field $\phi$,
named {\it quintessence} \cite{rp,Frieman,Zlatev}.
Cosmological observations indicate that
the potential of the quintessence field
should be very flat around the  present epoch. Consequently the
effective mass should be extremely small, $m_Q \sim 10^{-33}$ eV, which
surprisingly is also connected to the neutrino
masses {\it via} a see-saw formula $m_Q
\sim {m_\nu^2 / m_{pl}}$, with $m_{pl}$ being the Planck mass.

Are there any connections between the neutrinos and dark energy?
Given the arguments above it is quite interesting to make such
a speculation on this connection. If yes,
in terms of the language of particle physics, it requires the
existence of new dynamics and new
interactions between the neutrino and the dark energy sector. Recently
there are quite a few studies in the literature on the possible realization
of the models connecting neutrinos and dark
energy\footnote{In analog  to the idea of top quark condensate
as a mechanism of generating the electroweak scale, one might think that
the neutrino condensate
gives rise to the scale of dark energy. In this sense the dark energy
scalar field, such as quintessence behaves like a bound state of the neutrinos.
Effectively the system of dark energy sector consists of neutrinos and
(single or multi) scalars which interact with each other given
(approximately) by an
effective Lagrangian similar to Eqs.
(\ref{lag})
and (1).}\cite{hong1,gu,nelson,hong2,bi,kaplan,peccei,li,zhang,guen}.
Qualitatively these models have made at least two
interesting predictions:
1) neutrino masses are not constant, but vary during the evolution of
the universe; 2) CPT is violated in the neutrino sector due to the CPT
violating {\bf
Ether} during the evolution of the quintessence scalar field\cite{li}.
Quantitatively
these predictions will depend on the dynamics governing the coupled system
of the neutrinos and dark energy.

One of the possible couplings between the neutrinos and the
scalar field is the derivative interaction\cite{li}:
\begin{equation}
\label{nu0}
{\cal L} \sim \frac{{\partial_{\mu}}\phi}{\Lambda} {\bar \nu_L
\gamma^{\mu} \nu_L}\ .
\end{equation}
During the evolution of a homogeneous quintessence scalar field,
$\dot \phi$ does not vanish and gives rise to CPT violation in the
neutrino sector. However, since $\dot \phi$ is very
small at the present epoch this type of cosmological CPT
violation is predicted to be much
smaller than the current experimental limits.
But in the early universe with high
temperature it has been shown in Ref.\cite{li} that this CPT violation is
large
enough for the generation of the baryon number asymmetry
via leptogenesis. This new mechanism \cite{li,trodden}for
baryogenesis/leptogenesis provides a
unified picture for dark energy and baryon matter of the universe.

Another type of the interaction between the neutrinos and dark energy
is that the quintessence field couples to
the neutrino mass term. In the minimal extension of the
standard model of particle physics,
the neutrino masses can be described by a dimension-5 operator
  \be
\label{nu1}
 {\cal L}_{\not L}=\frac{2}{f}l_{L}l_{L}HH+h.c.\ ,
  \ee
  where $f$ is the scale of new physics beyond the standard model
  which generates the $B-L$ violation, $l_{L},  H$ are the
  left-handed lepton and Higgs doublets respectively. When the
  Higgs field gets a vacuum expectation value $<H> \sim v$, the
  left-handed neutrino receives a Majorana mass
  $m_{\nu} \sim \frac{v^{2}}{f}$. In Ref. \cite{gu} we considered
an interaction between the neutrinos and the
quintessence \be\label{coupl}
{\cal L}_{int}= \beta \frac { \phi }{M_{pl}} \frac{2 }{f} l_{L}l_{L}
H H+ h.c  , \ee
where $\beta $ is the coefficient characterizing the strength of the
interaction between quintessence and the neutrinos.
In this scenario the neutrino
masses vary during the evolution of the universe and
the neutrino mass limits imposed by the
baryogenesis are modified.

The operator (\ref{nu1})
is not renormalizable, which in principle can be generated by
integrating out the heavy particles. For example, in the model of
the minimal see-saw mechanism \cite{seesaw}for the neutrino masses, we
have
\be
{\cal L}_{\text{neutrino}}=h_{ij}\bar{l}_{Li}N_{Rj}H+
\frac{1}{2}M_{ij}\bar{N}^{c}_{Ri}N_{Rj}+h.c. ~,
\ee
 where $ M_{ij}$ is the mass matrix of
the right-handed neutrinos and the Dirac mass of neutrinos is given
by $m_{D}\equiv h_{ij} <H> $. Integrating out the heavy
right-handed neutrinos will generate the operator in (\ref{nu1}).
However, as pointed out in Ref. \cite{gu},
to make the light neutrino masses vary there are various
possibilities, such as by coupling the quintessence field to
 the Dirac masses and/or the Majorana masses of the right-handed
neutrinos. In Ref.\cite{bi} we have proposed a model
of mass varying right-handed neutrinos. In this model the
right-handed neutrino masses $M_i$ are assumed to be a
function of the quintessence field $M_i(\phi)=\overline{M}_i e^{\beta
\frac{\phi}{M_{pl}}}$. Integrating out the right-handed
neutrinos will generate a dimension-5 operator like in (3),
with the light neutrino masses varying in the following way
\be
\label{exp}
{\cal L}_{\not L}=e^{-\beta \frac{\phi}{M_{pl}}} \frac{2 }{f} l_{L}l_{L}
HH+ h.c.\; . \ee

In this paper we will give a systematic study on the effects for a
coupled system of light neutrinos and the quintessence field,
especially on the late-time evolution and the fate of the universe.
We will show numerically that 1) unlike the case in the absence of
the interaction the relic neutrino energy density $\Omega_\nu$ may not
decrease, instead it can track the energy density of the
quintessence field. In the future, all of the pressureless matter
may be diluted away and only the components of
the neutrinos and quintessence remain.
2) Neutrino masses will increase as the universe expands, however
when kinematically allowed the neutrinos will decay for instance
into electrons and pion mesons. Since the electrons and the mesons
are not coupled to the dark sector, the fate of the universe is
governed by the ``pure'' dynamics of the quintessence field. In this
paper we will quantify the moment when the neutrinos decay by the
numerical calculation.

The paper is organized as follows: in Section
II we will study the general features of
the dark energy models with mass
varying neutrinos and discuss analytically the behavior of the
evolution of the system. In Section III we present the numerical
results for some specific models.
Section IV gives discussions and conclusions.

\section{Dark Energy Models with Mass Varying Neutrinos}

In this section we will study in detail the
dark energy models where the neutrinos and the quintessence interact
with each other. Specifically we start with a class of models given by
the following Lagrangian
\be
\label{lag}
{\cal L}={\cal L}_\nu+{\cal L}_\phi+M(\phi)\bar{\nu}{\nu}\ ,
\ee
where
${\cal L}_\nu=\bar{\nu}i\partial \hspace{-0.2cm}/ \nu$ and
${\cal L}_\phi=\frac{1}{2}\partial_\mu \phi\partial^\mu\phi-V(\phi)$ with
$V(\phi)$ being the quintessence potential; $M(\phi)$ being the $\phi$-dependent
neutrino mass such as in Eqs. (\ref{coupl}) and (\ref{exp}).
In this paper we consider mainly the models of
neutrino coupling to the quintessence via the mass term.

As shown in Eq.(6) different dark energy models are specified by
the form of $V(\phi)$ and $M(\phi)$. In this paper we will
consider two classes of models, one has the exponential form of
$V(\phi)$ and $M(\phi)$ and the other has the power law form. In this
section we will give a general analysis of the behavior of
the system and its condition for the attractor solution.

The equation of motion for the scalar field $\phi$ is given by
\be\label{EM}
\ddot{\phi}+3H\dot{\phi}+\frac{dV}{d\phi}+\frac{dV_I}{d\phi}=0\ ,
\ee where \be \label{sour}
\frac{dV_I}{d\phi}=\frac{dM}{d\phi}n\left\langle\frac{M}{E}\right\rangle
\ee is the source term by the interaction between the neutrinos and
the scalar field, with $n$ and $E$ being the number density and energy of
the neutrinos respectively, $\langle \rangle$ indicates the thermal
average. For relativistic neutrinos, the term $\frac{dV_I}{d\phi}$
is greatly suppressed and the neutrinos decouple from quintessence.
For non-relativistic neutrinos, the effective potential of the
system is given by $V_{eff}(\phi)=V(\phi)+nM(\phi)$.

For this coupled system, the energy for each component does not
conserve. It is easy to get that the energy density of neutrinos
evolutes as \be \dot{\rho}_\nu+3H\rho_\nu=nM'\dot{\phi}~~, \ee where
$M'=dM/d\phi$. Correspondingly due to the conservation of
energy-momentum tensor of the
whole system, the fluid equation of the scalar field
$\phi$ is \be
\dot{\rho}_\phi+3H\rho_\phi(1+w_\phi)=-nM'\dot{\phi}\ , \ee where
$w_\phi=\frac{p_\phi}{\rho_\phi}
=\frac{\frac{1}{2}\dot{\phi}^2-V(\phi)}{\frac{1}{2}\dot{\phi}^2+V(\phi)}$
is the equation of state of the scalar field.
We can define the effective equation of states $w^{eff}_\nu$ and
$w^{eff}_\phi$ as \bea
\dot{\rho}_\nu+3H\rho_\nu(1+w^{eff}_\nu)=0\ ,\\
\dot{\rho}_\phi+3H\rho_\phi(1+w^{eff}_\phi)=0\ , \eea where \be
\label{wnu}
w^{eff}_\nu=-\frac{nM'\dot{\phi}}{3H\rho_\nu}=-\frac{1}{3}\frac{\partial\log
M}{\partial\phi}\phi' \ee and \be \label{wphi}
w^{eff}_\phi=w_\phi+\frac{1}{3}\frac{\partial\log
M}{\partial\phi}\phi'\frac{\rho_\nu}{\rho_\phi} \ee with
$\phi'=\frac{\partial\phi}{\partial \log a}=\frac{\dot{\phi}}{H}$.

The effective equation of states $w^{eff}_\nu$ and $w^{eff}_\phi$
describe the actual rate of the energy density decrease of the two
components respectively as the universe expands. When the two
equation of states become equal we expect the two components to
remain a constant ratio in energy densities. In the future when the
energy densities of all other components of the universe dilute away
and requiring $w^{eff}_\phi=w^{eff}_\nu$, we have
 \be
\Omega_\phi=\frac{\phi'^2}{3}+\frac{1}{3}\frac{\partial\log
M}{\partial\phi}\phi'\ , \ee where we have set $8\pi G=1$ and used
the relation $\Omega_\phi=\frac{\phi'^2}{6}+\Omega_V$. We can use
the requirement of a constant ratio of the energy densities to
derive the attractor solution of the system.

We first consider the class of models with exponential forms of
$V(\phi)$ and $M(\phi)$, which have an
attractor solution and will be shown below.
For $M(\phi)=\bar{M}e^{-\lambda\phi}$ we have
$\frac{\partial\log M}{\partial\phi}=-\lambda$. Requiring \be
\frac{d\Omega_\phi}{d\log a}=0 \ee we get \be
(2\phi'-\lambda)\phi''=0\ . \ee This leads to the solutions
$\phi'=\lambda/2$ or $\phi''=0$. The first solution gives
$\Omega_\phi < 0$. For $\phi''=0$, we have $\phi=
\phi_0+\lambda'\log a$. This solution leads to \be \label{om1}
\Omega_\phi=\frac{1}{3}(\lambda'^2-\lambda\lambda')\ . \ee For the
exponential potential $V=V_0e^{\beta \phi}$, we have the equation
of motion\cite{riotto} \be \label{om02}
\frac{\rho_m+\rho_\gamma+\rho_\nu+V}{1-\phi'^2/6}\frac{\phi''}{3}
+(\rho_m+\frac{2}{3}\rho_\gamma+2V)\frac{\phi'}{2}=-\beta
V+\lambda \rho_\nu\ . \ee From Eq. (\ref{om02}) we can get \be
\label{om2} \Omega_\phi= \frac{\beta
\frac{\lambda'^2}{6}+\lambda-\frac{\lambda^{\prime}}{2}
+\frac{\lambda'^3}{6}}{\frac{\lambda'}{2}+\beta+\lambda}\ . \ee
Eqs. (\ref{om1}) and (\ref{om2}) give the unique solution: \be
\lambda'=\frac{-3}{\lambda+\beta}\ . \ee This is the attractor
solution  given in Ref. \cite{riotto} in the discussions of a
coupled system of quintessence and dark matter.

The exponential potential has very good attractor
behavior and generically the coupled system is very close to the
attractor region today and leads to a constant ratio of
the two coupled components' energy densities.
This scenario seems to be less interesting to us,
since the evolution of the
universe is still dominated by the dynamics of the scalar field
with a negligible neutrino component.

In the following we will consider a class of power law models where
$M=\bar{M}\phi^{-\beta}$ and $V(\phi) = V_0 \phi^{\alpha}$ , $\beta$
and $\alpha$ are the two model parameters to be specified for the
numerical calculations in the next section.
 In this class of models,
$\frac{\partial\log M}{\partial \phi}=-\frac{\beta}{\phi}$ and we
have \be \Omega_\phi=\frac{\phi'^2}{3}-\frac{\beta\phi'}{3\phi}\ .
\ee Demanding that $\Omega_\phi$ remains a constant during the
evolution of the universe, one gets \be \label{equ}
2\phi^2\phi'\phi''-\beta\phi\phi''+\beta\phi'^2=0\ . \ee In the next
section we will present the detailed numerical results on the
evolutions of the universe. Specifically we will focus on the power
law models and on the behavior of the solution to the equation
above, when $\phi$ tends to zero and $\phi'/\phi$ tends to a
constant. Our results will show explicitly the
importance of the neutrino component and the interaction with quintessence in
determining  the evolution and the fate of the universe.

\section{Numerical Results}

In this section, with the power law forms of interaction and
potential, we will present our numerical results for the evolution
of the coupled system given in Eq. (\ref{lag}). We will show that
this class of models can satisfy the observations within the model
parameter space, furthermore predict different fates of the
universe. At high temperatures, the neutrinos are relativistic and
decoupled from the scalar field, as shown in Eq. (\ref{sour}). At
low temperatures, the neutrinos begin to affect the evolutions of
the scalar field. Specifically we notice that the system will
undergo a period of oscillation before entering the attractor
region. In the period of oscillation, the equation of state of
neutrino can be below $-1$, so its energy density can grow up and
become comparable with the scalar field or dark matter. Therefore,
the fate of the universe as we show below can be quite different
from the dark energy model in the absence of the interaction with
neutrinos. When  neutrinos do not interact with dark energy, the
density ratio between dark matter and neutrinos remains constant and
typically dark energy will dominant the late time universe. However
as shown below, the universe can be dominant with dark energy and
neutrinos after introducing the interaction,
with constant ratio of the energy densities of the two components.
On the other hand, the universe can also be dominated by neutrinos
and dark matter with negligible dark energy in another scenario. In
the numerical study we have taken the parameter values today as
$h \approx 0.7$,
$\Omega_\phi\approx 0.7$, $\Omega_m\approx 0.3$ and $\Omega_\nu <
0.02$.

\begin{figure}
\includegraphics[scale=.6]{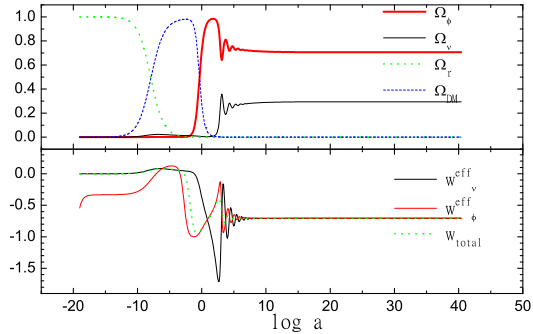}
\caption{\label{fig1}
The $\Omega_i= \frac{\rho_i}{\rho_c}$ and effective
equation of state $W$ as functions of the scale factor
$\log a$ for $V=V_0\phi^{5}$ and $M=\bar{M}\phi^{-12}$.
}
\end{figure}

In Fig. \ref{fig1}, by specifying  the potential
$V=V_0\phi^{5}$ and the neutrino mass $M=\bar{M}\phi^{-12}$, we plot the evolution of $\Omega_i=
\frac{\rho_i}{\rho_c}$ and the effective equation of states as
function of the scale factor $\log a$. The
effective equation of states in the figure are defined in Eqs.
(\ref{wnu}) and (\ref{wphi}). $w_{total}=P_{total}/\rho_{total}$ here is the equation of state of the
whole system.  After a short period of oscillation, the energy density fraction of the scalar field decreases
while the component of neutrino increases to around 30\%. The
universe will be dominated by the two components in the future and
the ratio of the two components keeps constant for ever. This
picture corresponds to a solution of the forementioned equation (\ref{equ})
when $\phi$ approaches zero and $\phi'/\phi$ remains
nonvanishing.\footnote{For simplicity we have taken neutrinos to be
matter like in the early epoch. This does not
change our late time picture when considering the neutrino
behavior at the early times.}

\begin{figure}
\includegraphics[scale=.6]{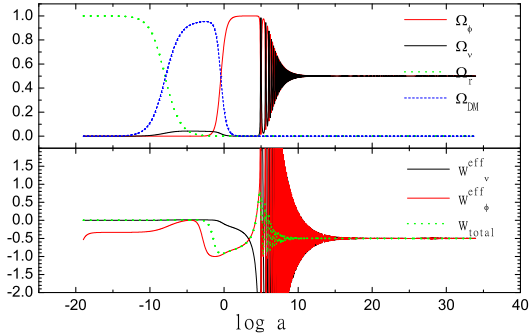}
\caption{\label{fig2}
The $\Omega_i$ and the effective equation of state $W$
as functions of the scale factor
$\log a$ for $V=V_0\phi^{4}$ and $M=\bar{M}\phi^{-4}$.
}
\end{figure}

\begin{figure}
\includegraphics[scale=.6]{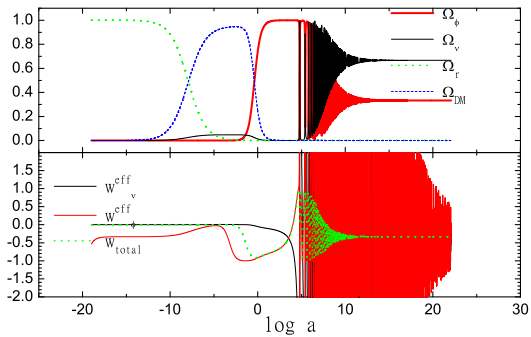}
\caption{\label{fig3}
The $\Omega$ and the effective equation of state $W$ as
functions of the scale factor
$\log a$ for $V=V_0\phi^{4}$ and $M=\bar{M}\phi^{-2}$.
}
\end{figure}

In the case of interacting neutrino and dark energy there is
typically a period of ``build-in'' phase before the system arrives
the late time stable states. Fig. \ref{fig1} stands for the case
where the system enters quickly the stable phase. We find that to
achieve a universe with a larger fraction of neutrinos,
it will typically take a
considerably longer time for the ``build-in'' phase. As an
example we consider a model with $V=V_0\phi^{4}$ and the neutrino mass
$M=\bar{M}\phi^{-4}$. In Fig. \ref{fig2} we plot the evolution of
$\Omega_i$ and the effective equation of states for this model.
One can see after the period of oscillation, the universe approaches
to a state consisting equally of the neutrino and the scalar field:
$\Omega_\nu=\Omega_\phi=50\%$. The equation of state of the system
at the tracking region is $w_{total}=-0.5$. A more interesting
case is plotted in Fig. \ref{fig3} for $V=V_0\phi^{4}$ and
$M=\bar{M}\phi^{-2}$. The universe will finally be dominated by the neutrinos
with $\Omega_\nu\approx 2/3$ while $\Omega_\phi\approx 1/3$. The
equation of state of the system is  $w_{total}\approx -1/3$.

\begin{figure}
\includegraphics[scale=.6]{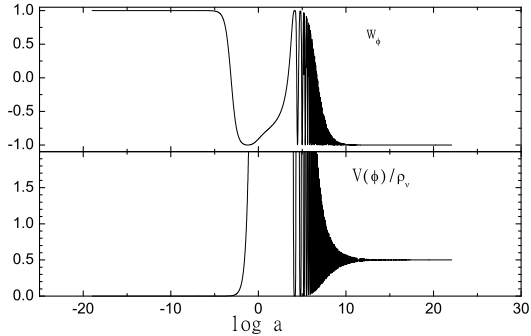}
\caption{\label{fig3x} $w_{\phi}$ and the ratio of $V(\phi)$ with
neutrino energy density as functions of the scale factor $\log a$
for $V=V_0\phi^{4}$ and $M=\bar{M}\phi^{-2}$. }
\end{figure}

The oscillating phase is determined by two conditions. One is the
asymptotic behavior of $V'_{eff}$. In the case of a power law
quintessence potential and neutrino masses, $V'_{eff}$ gets
asymptotically to zero during the evolution. The other condition is
the magnitude of $V''$ compared with the Hubble expansion rate. The
scalar field starts to oscillate at the time $V''_{eff}\sim H^2$. Due to the
Hubble friction the oscillation damps with time. When the
oscillating amplitude is small enough the kinetic  term of $\phi$
becomes negligible, i.e. $\frac{1}{2}\dot{\phi}^2 \ll V(\phi)$, and
$w_{\phi}$ tends to $-1$. For the power law form we will then have
$\alpha V(\phi)\simeq \beta \rho_{\nu}$,
$w_{total}=\frac{\frac{1}{2}\dot{\phi}^2-V(\phi)}{\frac{1}{2}\dot{\phi}^2+V(\phi)+\rho_{\nu}}
\simeq \frac{-V(\phi)}{V(\phi)+\rho_{\nu}}=\frac{-\beta}{\alpha+
\beta }$ and $\Omega_{\phi}\simeq \frac{\beta}{\alpha+ \beta }$.
Taking the model in Fig. \ref{fig3} as an example, we delineate in
Fig. \ref{fig3x} the corresponding $w_{\phi}$ and $V(\phi)/
\rho_{\nu}$ as functions of the scale factor $\log a$. The
contribution of the kinetic term oscillates and decreases with time.
After the freezing of $w_{\phi}=-1$, the ratio $V(\phi)/ \rho_{\nu}$
approaches $1/2$ and $V'_{eff}\sim 0$. Comparing with Fig.
\ref{fig3}, we notice that $w_{total}$ and $\Omega_{\phi}$ become
nearly constant after $w_{\phi}$ approaches -1. However, we should
point out that the damped oscillation lasts for a longer time and
 $\Omega'_{\phi}/\Omega_{\phi}$ is not yet
negligible for $\log a > 10$ in the shown range of $\log a$ in
Fig. \ref{fig3}
and Fig. \ref{fig3x}. Meanwhile the two effective equations of state
remain damped oscillating around $w_{total}$.

\begin{figure}
\includegraphics[scale=.6]{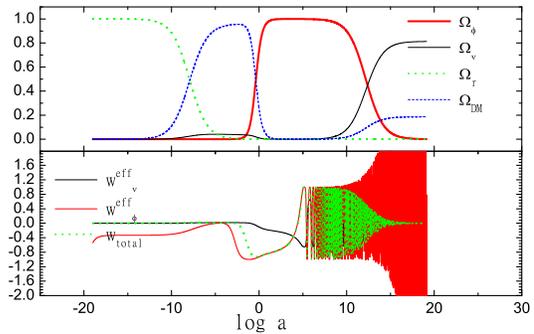}
\caption{\label{fig4} The $\Omega$ and the effective equation of
state $W$ as functions of the scale factor $\log a$ for
$V=V_0\phi^{4}$ and $M=\bar{M}e^{-4\phi/M_{Pl}}$. }
\end{figure}

As shown above in the class of models
with the scalar potential $V(\phi)$ and the neutrino mass
$M(\phi)$ being the power law form the
neutrino masses keep increasing and, unlike the
dark matter, its energy density does not dilute away. In some cases
neutrinos will even dominate
over the quintessence. These results, however depend on the
couplings of the neutrinos to quintessence. As an example we take
$V(\phi)=V_0\phi^{\alpha}$ while $M=\bar{M}\exp(-\beta \phi)$. The
behavior of neutrinos and dark energy cannot remain symmetric and
their ratio cannot take as
constant during late time evolution. The dark
energy will transfer its energy density to neutrinos before their
decoupling. The universe becomes neutrino dominant with a
constant ratio between $\Omega_{\phi}$ and $\Omega_m$. In Fig. \ref{fig4} we
realize this picture with $V(\phi)=V_0\phi^4$ and
$M=\bar{M}\exp(-4 \phi/M_{Pl})$. The universe evolves approaching a
state dominated by matter and neutrino, where the effective
equation of state $w_{total}$ is zero.
Due to the different behavior of $V'_{eff}$
from the previous power law cases, the neutrino mass in Fig. \ref{fig4} will
become constant in the future when the universe stops
acceleration.


\begin{figure}
\includegraphics[scale=.6]{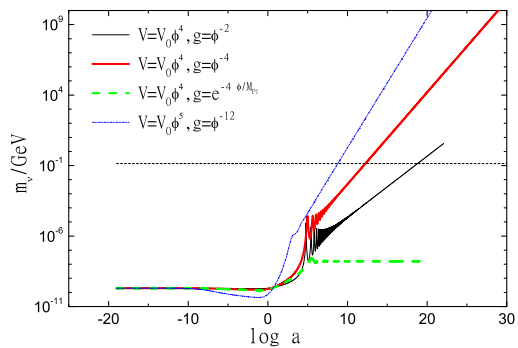}
\caption{\label{fig5} The neutrino mass as functions of the scale
factor $\log a$, where $g= M(\phi)/\bar{M}$. }
\end{figure}

Generally speaking the mass of neutrinos will vary as the evolution
of the coupled system. It is easy to show that in general
$M(\phi)=\bar{M} a^{-3 w ^{eff}_\nu}$ with $a$ being the scale
factor. For $w^{eff}_\nu < 0$, the neutrinos are lighter in the
past, but become heavier in the future. In Fig. \ref{fig5}, we plot
the neutrino mass as a function of the scale factor $\log a$ for the
models discussed in this section. It should be noticed that the
neutrino mass will increase forever as the expansion of the universe
for the power law cases. When the neutrinos are heavy enough,
however they will decay to electron and pion $\nu\to e^\pm+\pi^\mp$.
The life time of the decay is given by \be \tau_\nu= 6.\times
10^{-9}\ \text{sec}\ \ \cdot \left( 1-\frac{m_\pi^2}{m_\nu^2}
\right)^{-2}\ . \ee Therefore, the neutrinos will decay away
immediately once they cross the energy threshold. The energy density
of neutrinos will then be transferred to matter, which may
eventually be diluted away with the expansion of the universe. In
this case the scalar field will finally dominate the universe and
determine the evolution of the universe. In Fig. \ref{fig5} for an
intuitive view, the moment when the decay occurs is set as
$m_\nu=m_\pi+m_e$.

\section{Discussions  and Conclusions}

In this paper we have studied the dynamics of the
coupled system described by Eq. (\ref{lag}). We  paid particular
attention to the effects of the back reaction of the thermal
neutrino
bath in the universe to the evolution of the scalar field
--- the quintessence ---
due to the $\phi$-dependent mass term\cite{bran}.

The coupled system has many interesting phenomenologies. First of
all, during the evolution of the universe the neutrino  masses vary.
This effect will make the matter power spectrum observed from the
Large Scale Structure(LSS) survey different from that for a constant
neutrino mass. Consequently the present cosmological
 limits on the neutrino mass
$m_i < 0.6 eV$ will be relaxed in the scenario with mass varying
neutrinos \cite{bin}. In the current paper we mainly focus on the
models where neutrinos can track dark energy. Although the mass of
neutrinos varies significantly around the tracking regime, it has
changed very little around today. This is due to the fact that the
coupled system of neutrinos and dark energy
has been stringently constrained by
the current observations. It is trivial if the system has entered
the tracking regime today when the fraction of the neutrino
energy density is negligible\cite{caution}. Meanwhile as the tracking
regime has been kept in the {\bf far} future from today, the mass of
neutrinos has changed little till the present epoch due to the slow
rolling of $\phi$. In this sense our scenario is hard to distinguish
experimentally from the standard case with constant neutrino masses.
On the other hand, however,
if we do not expect a neutrino dominant future universe and the
present epoch is closer to the tracking regime, neutrino mass could
have evolved from the past to now, as shown in the example of
$V=V_0\phi^{5}$ and $M=\bar{M}\phi^{-12}$ in Fig. \ref{fig5}. We
could then expect some distinguishable imprints from the uncoupled
case\cite{bin}.

In addition,
if the neutrino mass is produced through the seesaw mechanism and the
scalar field only couples with the right-handed neutrinos,
this scenario will change the predictions of the leptogenesis
mechanism. As shown explicitly in Ref. \cite{bi} the reheating temperature
required by leptogenesis can be lowered
and the bound on the light neutrino mass set
by leptogenesis can also be relaxed.

Finally, the derivative coupling in Eq. (\ref{nu0}) contributes also to
the kinetic term. It is easy to show that
the quantum effect at zero temperature vanishes.
However, at finite temperature $T$ the corrections are given by
\be
{\cal L}_{correct}^T=\frac{1}{2}\left( \frac{T^2}{12\Lambda^2}\right)
\partial_\mu \phi\partial^\mu\phi \ .
\ee The effect is negligible at low temperature $T\ll \Lambda$.
However, it could be important at very high temperatures when $T\sim
\Lambda$. Ref.\cite{li} has also considered another type of the back
reaction caused by the derivative couplings in Eq. (\ref{nu0}) and
shown explicitly the negligible effects at low temperature
\footnote{For the phantom scalar field, the effective Lagrangian
after including the quantum effects from the derivative couplings is
given by \be {\cal L}_{eff}=\frac{1}{2}\left(
-1+\frac{T^2}{12\Lambda^2}\right)
\partial_\mu \phi\partial^\mu\phi \ .
\ee
At high temperatures the system is a canonical scalar field while
at low temperature a `phase transition' occurs and the system
turns to the phase of phantom. This transition is somewhat favored
by the present SN Ia observations \cite{obs}.
}.


We should point out that the main purpose of the current paper is
the phenomenological study of the coupled system of dark energy and
neutrinos. We do not try to build the dynamical relation of
neutrinos and dark energy and a possibility of neutrino condensate
to achieve dark energy is pointed out in the footnote 1.  Attempts
for a possible dynamical link between the values of the neutrino
mass and dark energy today have been studied in Refs.
\cite{Frieman,fhsw}. Furthermore as shown in the previous studies of
tracking dark energy models one usually needs some fine tuning for
the energy density of quintessence today\cite{Zlatev}.

In summary we have presented a systematic study on the
cosmological evolution of the dark energy model with a coupled
system including a dynamical scalar field (the quintessence) and the
neutrinos. We have shown that the dynamics of this system drives the
universe accelerating at present and leads to various interesting
possibilities on the evolution of the universe. In our models, the
neutrinos may dominate the universe in the future. However, in most
scenarios, the neutrinos will eventually decay away.

\begin{acknowledgments}
We thank Peihong Gu, Roberto Peccei, Yun-Song Piao and Bing-Lin
Young for discussions. This work is supported in part by the Natural
Science Foundation of China under the Grant Nos. 10575111, 10105004,
10120130794, 90303004,
19925523  and by the Ministry of Science and Technology of China
under the Grant No. NKBRSF G19990754.
\end{acknowledgments}

\newcommand\AJ[3]{~Astron. J.{\bf ~#1}, #2~(#3)}
\newcommand\APJ[3]{~Astrophys. J.{\bf ~#1}, #2~ (#3)}
\newcommand\APJL[3]{~Astrophys. J. Lett. {\bf ~#1}, L#2~(#3)}
\newcommand\APP[3]{~Astropart. Phys. {\bf ~#1}, #2~(#3)}
\newcommand\CQG[3]{~Class. Quant. Grav.{\bf ~#1}, #2~(#3)}
\newcommand\JETPL[3]{~JETP. Lett.{\bf ~#1}, #2~(#3)}
\newcommand\MNRAS[3]{~Mon. Not. R. Astron. Soc.{\bf ~#1}, #2~(#3)}
\newcommand\MPLA[3]{~Mod. Phys. Lett. A{\bf ~#1}, #2~(#3)}
\newcommand\NAT[3]{~Nature{\bf ~#1}, #2~(#3)}
\newcommand\NPB[3]{~Nucl. Phys. B{\bf ~#1}, #2~(#3)}
\newcommand\PLB[3]{~Phys. Lett. B{\bf ~#1}, #2~(#3)}
\newcommand\PR[3]{~Phys. Rev.{\bf ~#1}, #2~(#3)}
\newcommand\PRL[3]{~Phys. Rev. Lett.{\bf ~#1}, #2~(#3)}
\newcommand\PRD[3]{~Phys. Rev. D{\bf ~#1}, #2~(#3)}
\newcommand\PROG[3]{~Prog. Theor. Phys.{\bf ~#1}, #2~(#3)}
\newcommand\PRPT[3]{~Phys.Rept.{\bf ~#1}, #2~(#3)}
\newcommand\RMP[3]{~Rev. Mod. Phys.{\bf ~#1}, #2~(#3)}
\newcommand\SCI[3]{~Science{\bf ~#1}, #2~(#3)}
\newcommand\SAL[3]{~Sov. Astron. Lett{\bf ~#1}, #2~(#3)}

\end{document}